\documentclass{iaus}
\usepackage{graphicx}

\title[S266.~~The validity of observed cluster masses and ages] 
{The range of validity of cluster masses and ages derived from broad-band photometry}

\author[J. Ma\'{\i}z Apell\'aniz]   
{J. Ma\'{\i}z Apell\'aniz$^1$}

\affiliation{$^1$Instituto de Astrof\'{\i}sica de Andaluc\'{\i}a-CSIC, Granada, Spain\\
             email: {\tt jmaiz@iaa.es}}

\pubyear{2009}
\volume{266}  
\pagerange{1--6}
\jname{Star Clusters}
\editors{Richard de Grijs \& Jacques R. D. L\'epine, eds.}
\begin{document}

\maketitle

\begin{abstract}
I analyze the stochastic effects introduced by the sampling of the stellar initial mass function 
(SIMF) in the derivation of the individual masses and the cluster mass function (CMF) from 
broad-band visible-NIR unresolved photometry. The classical method of using unweighted $UBV$ 
photometry to simultaneously establish ages and extinctions of stellar clusters is found to be 
unreliable for clusters older than $\approx 30$ Ma, even for relatively large cluster masses. 
On the other hand, augmenting the filter set to include longer-wavelength filters and using 
weights for each filter increases the range of masses and ages that can be accurately measured 
with unresolved photometry. Nevertheless, a relatively large range of masses and ages is found 
to be dominated by SIMF sampling effects that render the observed masses useless, even when 
using $UBVRIJHK$ photometry.
\end{abstract}

\firstsection 
\section{Description}

This work is the third of a series in which we are analyzing the possible biases present in mass functions
(\cite{MaizUbed05}; \cite{Maiz08}). I have used a combination of analytical approximations and Monte Carlo simulations 
to study the effect of the stochastic sampling of the stellar IMF (SIMF, assumed to be Kroupa) in the determination of 
masses and ages of unresolved stellar custers from broad-band photometry. For this purpose I have used the new (3.1)
version of CHORIZOS (\cite{Maiz04}), which incorporates an evolutionary synthesis module. For a given mass and age,
I generated a minimum of 10\,000 realizations of the SIMF using solar metallicity [a] Geneva isochrones for massive stars
and [b] Padova isochrones with new AGB treatment for low and intermediate masses. In each case I measured the mass (and
age, if appropriate) assuming that the cluster has a well-sampled SIMF. From the individual realizations I derived [a]
the observed cluster mass and age distributions for clusters of a (fixed) real mass and age and [b] the observed 
cluster mass functions (CMFs) for a real truncated power-law CMF with a slope of $\gamma=-2.0$. Three different cases 
in order of complexity were considered: [a] single-filter observations of clusters with known age and extinction, [b]
multi-filter observations of clusters with unknown age and known extinction, and [c] multi-filter observations of 
clusters with unknown age and extinction. In the first case masses were computed directly by converting filter-convolved
luminosities to masses while in the second and third cases a $\chi^2$ minimization code (CHORIZOS) was used to derive the
masses and ages. The reader is referred to \cite{Maiz09} for further details.

\section{Case 1: Single filter, known age and extinction}

\begin{figure}
\centerline{\includegraphics*[width=\linewidth, bb=130 400 475 720]{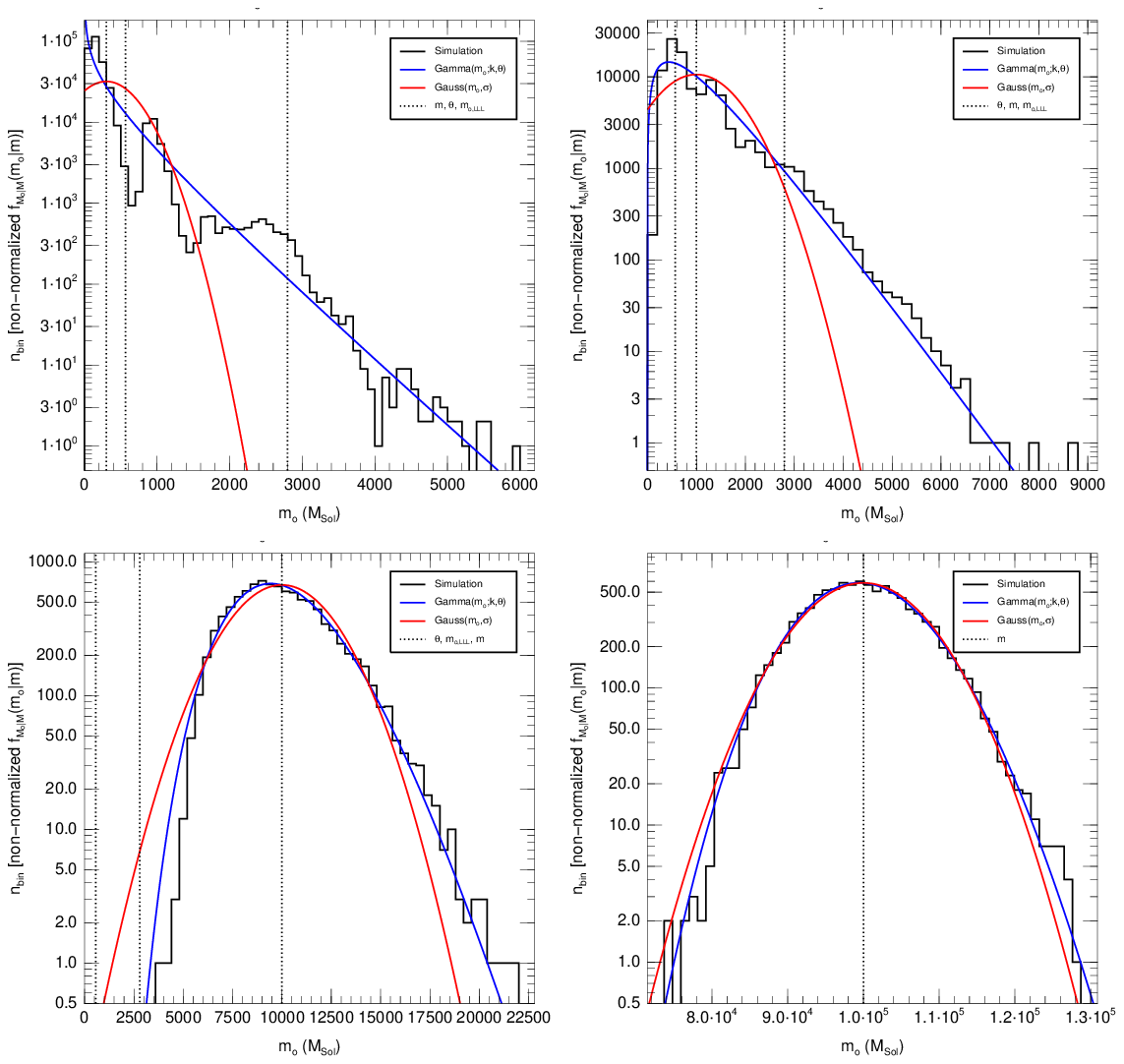}}
\caption{Results of the Monte Carlo simulations of $f_{M_{\rm o}|M}(m_{\rm o}|m)$ and Gaussian 
and Gamma fits for 10 Ma clusters observed in $V$ with four real masses: 300~M$_\odot$
(upper left), 1000~M$_\odot$ (upper right), 10\,000~M$_\odot$ (lower left), and 
100\,000~M$_\odot$ (lower right). The Poisson mass $\theta$ for this filter and age is 
566 M$_\odot$. For large real masses, both fits provide reasonable approximations but for small 
values of $m$, the Gamma approximation is significantly better. The vertical dotted lines show 
the values of the real mass $m$ and, when visible, of $\theta$ and $m_{\rm o,LLL}$ (the observed 
cluster mass obtained when one mistakes the brightest possible star in the isochrone for a 
cluster).}
\label{fmocm}
\end{figure}	

\begin{figure}
\centerline{\includegraphics*[width=\linewidth]{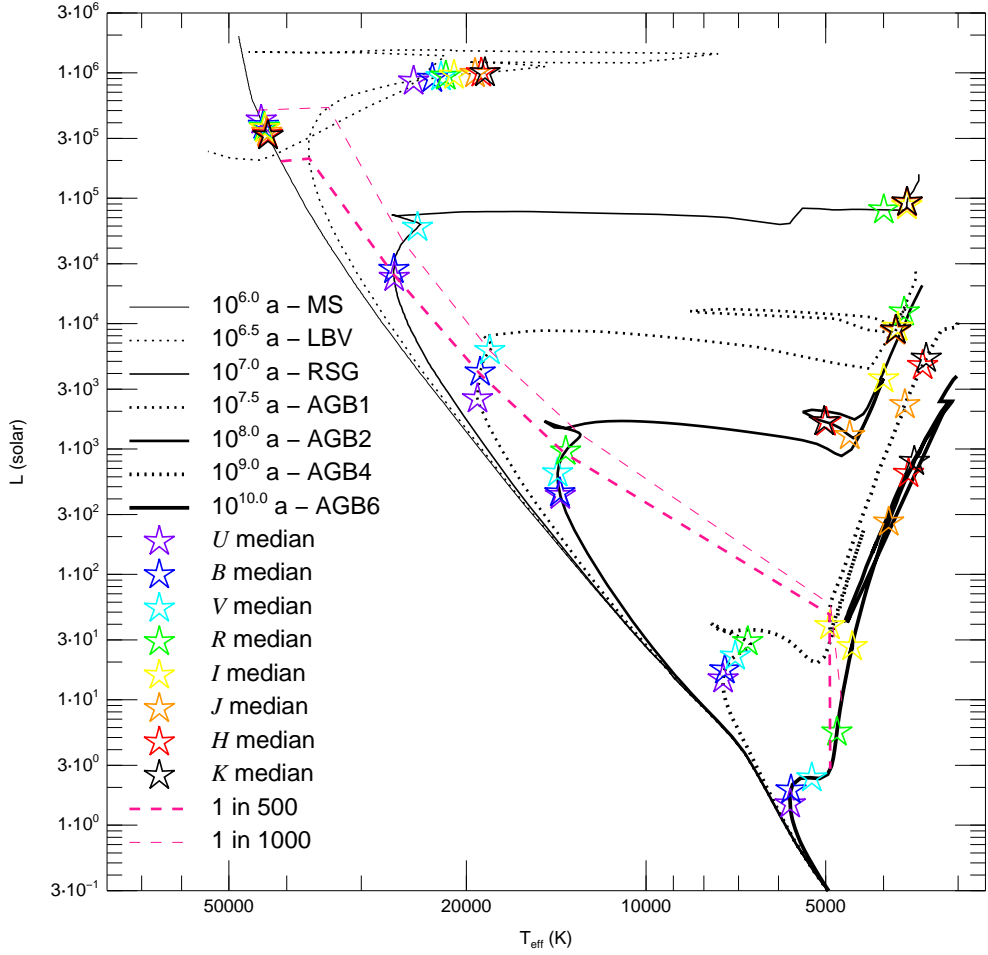}}
\caption{Isochrones for seven of the nine ages used in this work. For each isochrone, symbols of 
different colors are used to indicate the median mass for the luminosity in the eight filters $U$, 
$B$, $V$, $R$, $I$, $J$, $H$, and $K$ assuming a Kroupa IMF between 0.1~M$_\odot$ and 
120~M$_\odot$. The two dashed lines join the points in each isochrone beyond which 0.2\% and 0.1\%, 
respectively, of the remaining stars in a well-sampled Kroupa IMF are located. The two remaining 
isochrones are not included for the sake of clarity.}
\label{isoc}
\end{figure}	

\begin{table}
\begin{center}
\caption{Values of $\theta$ (in M$_\odot$) for different ages and filters for clusters 
observed with a single filter and known age and extinction. The labels below each age 
indicate the type of the brightest possible star.
For a given filter, $\theta$ starts at a low value in 
the MS phase, rapidly grows in the LBV phase, decreases during the RSG and 
early AGB phases, and then experiences a moderate growth when the cluster becomes old. For a 
given age, $\theta$ increases as a function of wavelength with the only exception of the MS 
phase, where the effect is the opposite (but rather weak).}
\begin{tabular}{crrrrrrrrr}
\hline
       &    1 Ma & 3.16 Ma &   10 Ma & 31.6 Ma &  100 Ma &  316 Ma &    1 Ga & 3.16 Ga &   10 Ga \\
Filter &      MS &     LBV &     RSG &    AGB1 &    AGB2 &    AGB3 &    AGB4 &    AGB5 &    AGB6 \\
\hline
$U$    &     235 &     945 &     160 &     149 &      53 &      33 &      23 &      26 &      42 \\
$B$    &     200 &  1\,314 &     192 &     305 &      51 &      32 &      27 &      52 &      92 \\
$V$    &     190 &  1\,555 &     566 &     370 &      89 &      38 &      51 &     106 &     163 \\
$R$    &     186 &  1\,722 &  1\,034 &     443 &     151 &      72 &     109 &     173 &     242 \\
$I$    &     178 &  1\,949 &  1\,585 &     616 &     273 &     463 &     476 &     477 &     581 \\
$J$    &     157 &  2\,741 &  2\,240 &  1\,170 &     708 &  3\,598 &  2\,930 &  2\,702 &  2\,445 \\
$H$    &     146 &  3\,081 &  2\,476 &  1\,640 &  1\,281 &  5\,184 &  4\,828 &  4\,499 &  4\,123 \\
$K$    &     140 &  3\,333 &  2\,542 &  1\,771 &  1\,527 &  5\,717 &  5\,688 &  5\,512 &  5\,058 \\
\hline
\end{tabular}
\end{center}
\label{theta}
\end{table}

\begin{figure}
\centerline{\includegraphics*[width=0.47\linewidth]{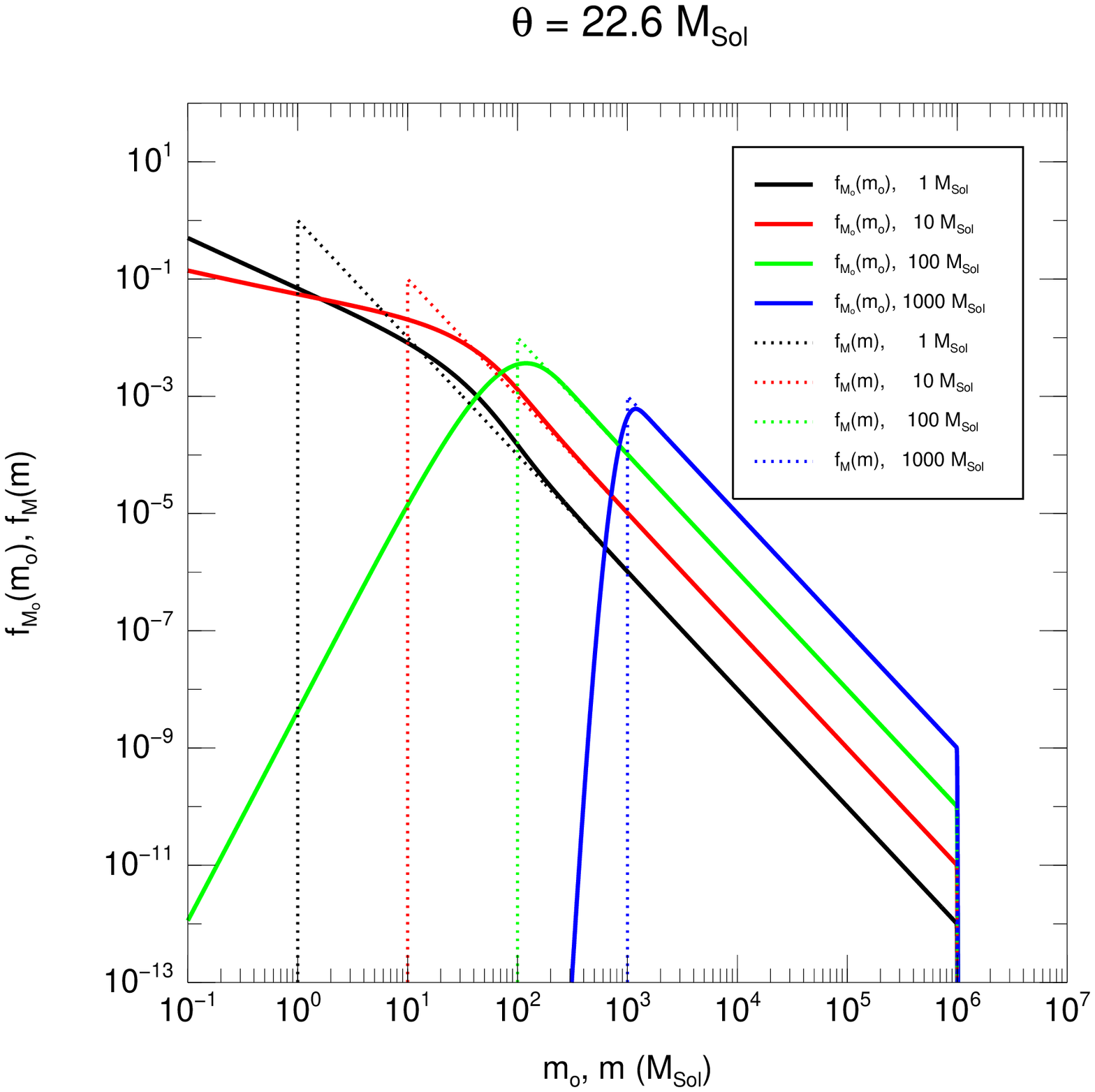}
            \includegraphics*[width=0.47\linewidth]{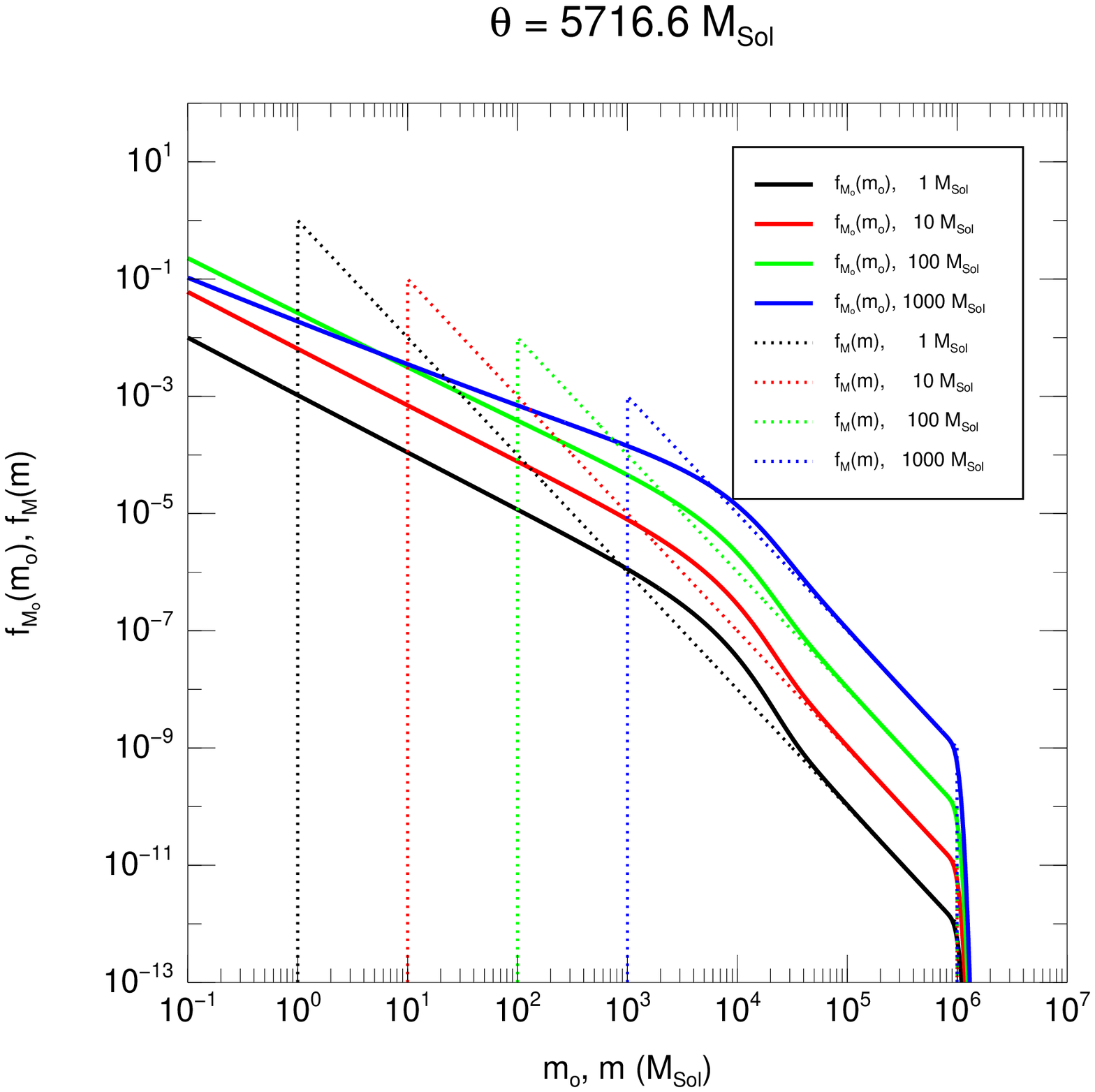}}
\caption{Observed (continuous lines) and real (dotted lines) CMFs for the minimum and maximum
values of $\theta$ in Table~\ref{theta}.
The left panel corresponds to 1 Ga clusters observed with $U$ and the right panel 
to 316 Ma clusters observed with $K$. Each panel shows four cases of the lower mass cutoff for 
$f_M(m)$ from 1~M$_\odot$ to 1000 M$_\odot$. As $\theta$ increases, the hump becomes more pronounced 
and moves towards the right.}
\label{fmofm}
\end{figure}	

\begin{itemize}
  \item For a given cluster mass ($m$), the observed cluster mass ($m_{\rm o}$) distribution can be approximated by a 
        Gamma function better than by a Gaussian distribution (Figure~\ref{fmocm}):
\begin{equation}
f_{M_{\rm o}|M}(m_{\rm o}|m) = A m_{\rm o}^{m/\theta-1} e^{-m_{\rm o}/\theta} \theta^{-m/\theta}.
\end{equation}
  \item The  Poisson mass $\theta$ is a measurement of the stochasticity and (strongly) depends on the age and the 
        filter used to derive $m_{\rm o}$ (Table~\ref{theta}).
\begin{equation}
\overline{m_{\rm o}} = m \;\;\;\; ; \;\;\;\;  \sigma_{m_{\rm o}} = \sqrt{\theta m}.
\end{equation}
  \item The large differences in $\theta$ are a consequence of the different fraction of stars located above the 
        median mass for the luminosity in each filter for a given isochrone (Figure~\ref{isoc}).
  \item The observed CMF, $f_{M_{\rm o}}(m_{\rm o})$, differs from the real (truncated power-law) CMF, $f_M(m)$, in two 
        aspects (Figure~\ref{fmofm}): [a] For large masses, $f_{M_{\rm o}}(m_{\rm o})$ is similar to $f_M(m)$ but, as the 
	cluster mass decreases, at one point a "hump" or overdensity appears. [b] To the left of the hump,
	$f_{M_{\rm o}}(m_{\rm o})$ asymptotically approaches a power law with a slope larger than $-1.0$, with some 
	clusters having $m_{\rm o}$ values lower than the real cutoff in $m$.
  \item Using an analytical approximation, it can be shown show that if one is willing to tolerate a systematic error in 
        the CMF slope $\gamma$ up to $\Delta\gamma$, then one can only analyze observed masses:
\begin{equation}
m_{\rm o} > 0.5\gamma(\gamma+1)\theta/\Delta\gamma,
\end{equation}
        e.g. for $\gamma = -2.0$ and $\Delta\gamma = 0.1$, $m_{\rm o} > 10\, \theta$.
\end{itemize}

\section{Case 2: Multiple filters, unknown age, and known extinction}

\begin{figure}
\centerline{\includegraphics*[width=0.42\linewidth]{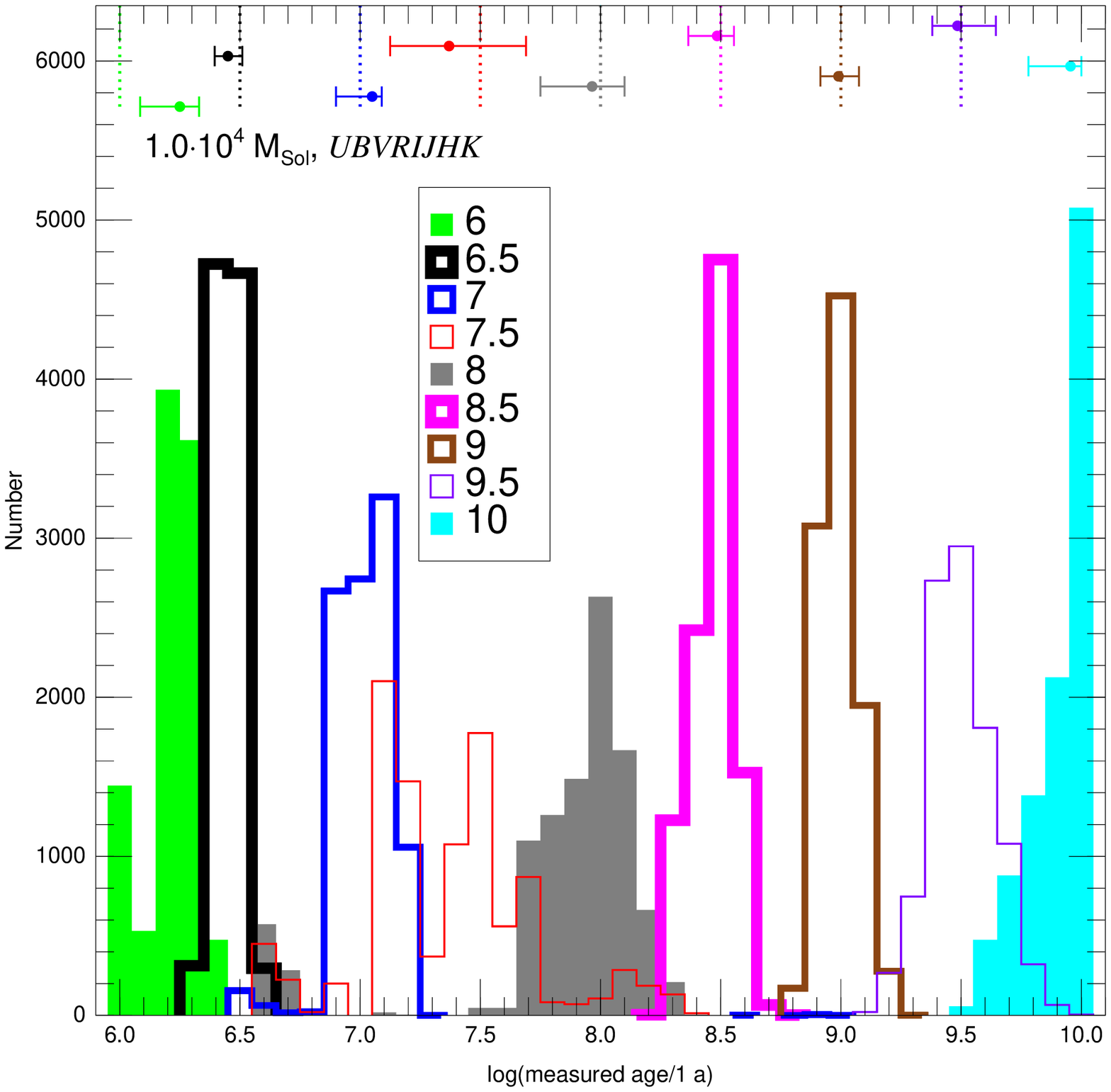}
            \includegraphics*[width=0.42\linewidth]{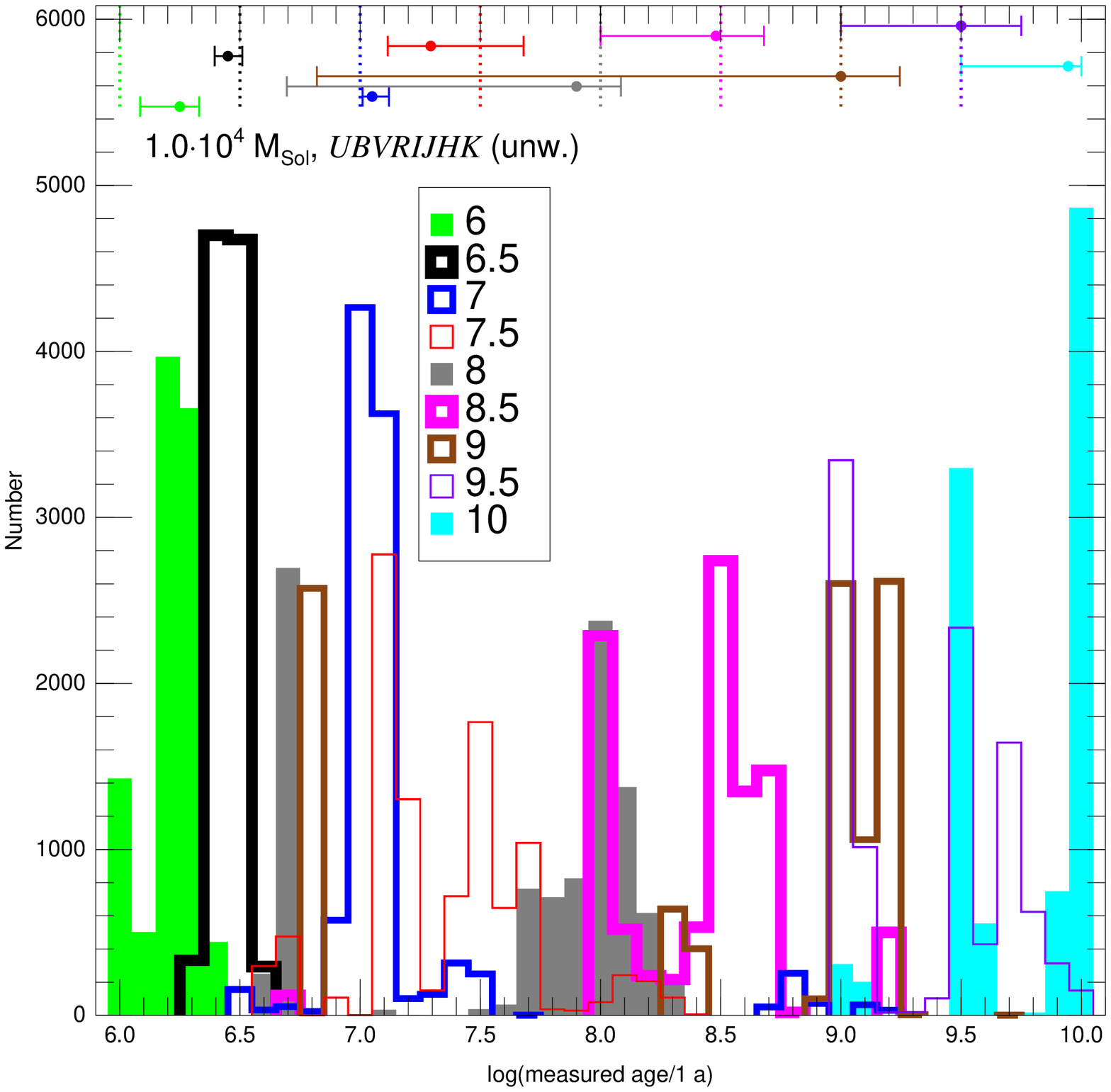}}
\centerline{\includegraphics*[width=0.42\linewidth]{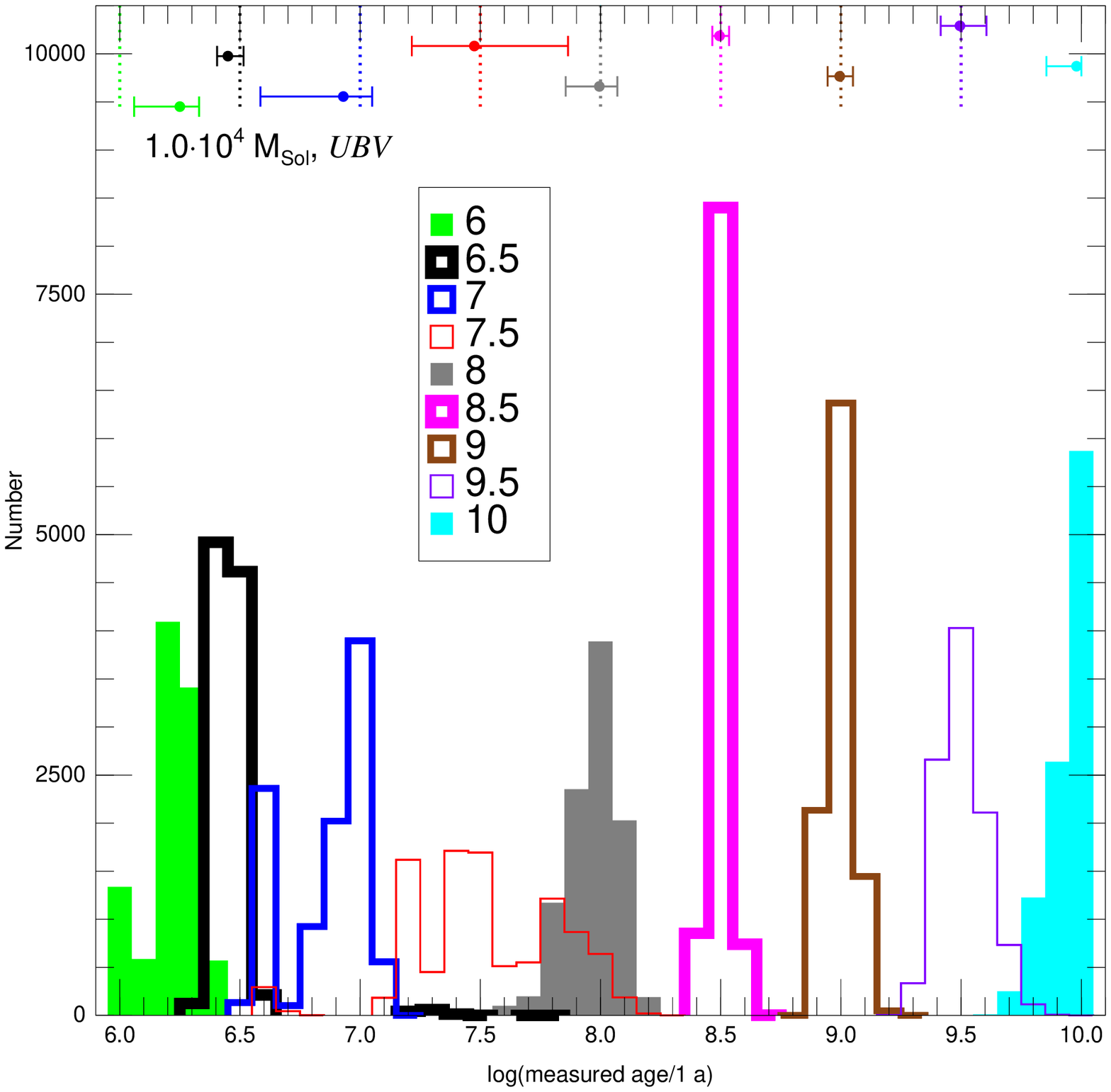}}
\caption{Distribution of observed ages for the nine input ages and three execution types (1: upper left, 
2: upper right, 3: bottom) for clusters of $10^4$ M$_\odot$ and unknown ages. At the top of each panel 
the dashed line indicates the real age while the symbols and error bars provide the median and inferior 
and superior uncertainties (1-sigma equivalents) of each distribution.}
\label{agedist1}
\end{figure}	

\renewcommand{\labelenumi}{\arabic{enumi}.}

	I used three execution types:

\begin{enumerate}
  \item $UBVRIJHK$ photometry with single-band $\theta$-dependent weights for each filter.
  \item $UBVRIJHK$ photometry with constant weights.
  \item $UBV$ photometry with single-band $\theta$-dependent weights for each filter.
\end{enumerate}

	The observed age distributions for $10^4$ M$_\odot$ clusters are shown in Figure~\ref{agedist1}. The main results
are:

\begin{itemize}
  \item The ages calculated with $UBVRIJHK$ photometry and constant weights (i.e. the standard method) are highly 
        uncertain even for $10^4$ M$_\odot$ clusters. 
  \item Using weights based on the stochasticity of each filter significantly narrows the observed-age distribution for 
        $10^4$ M$_\odot$ clusters, thus allowing for the obtention of ages with relatively small uncertainties.
  \item $UBV$ and $UBVRIJHK$ photometry provide relatively similar results, with the former having the advantage for old 
        clusters and the latter for intermediate-age ones. 
  \item The AGB1 stage (31.6 Ma) has significantly larger uncertainties than the rest.
\end{itemize}

\section{Case 3: Multiple filters, unknown age and extinction}

\begin{figure}
\centerline{\includegraphics*[width=0.42\linewidth]{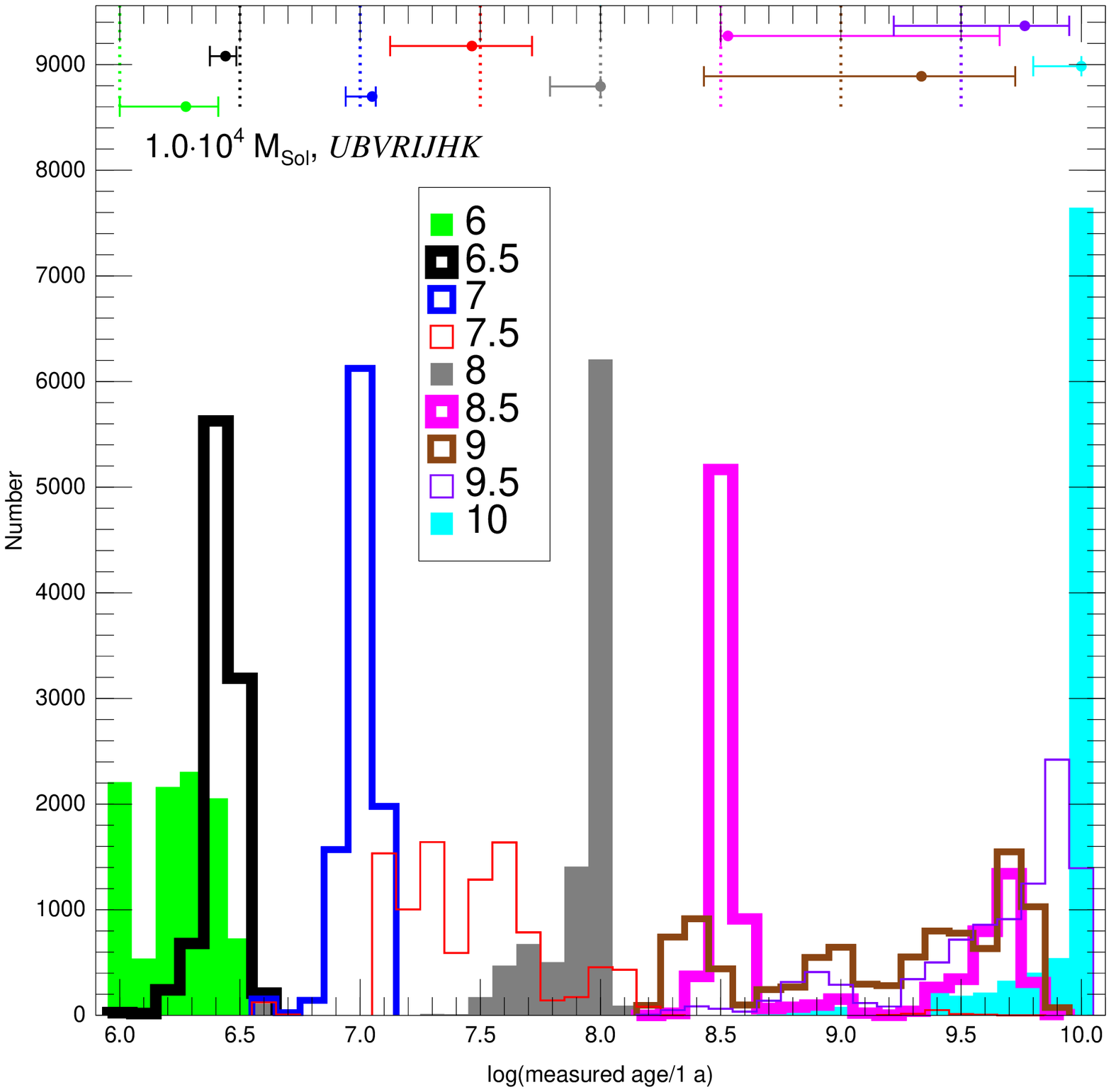}
            \includegraphics*[width=0.42\linewidth]{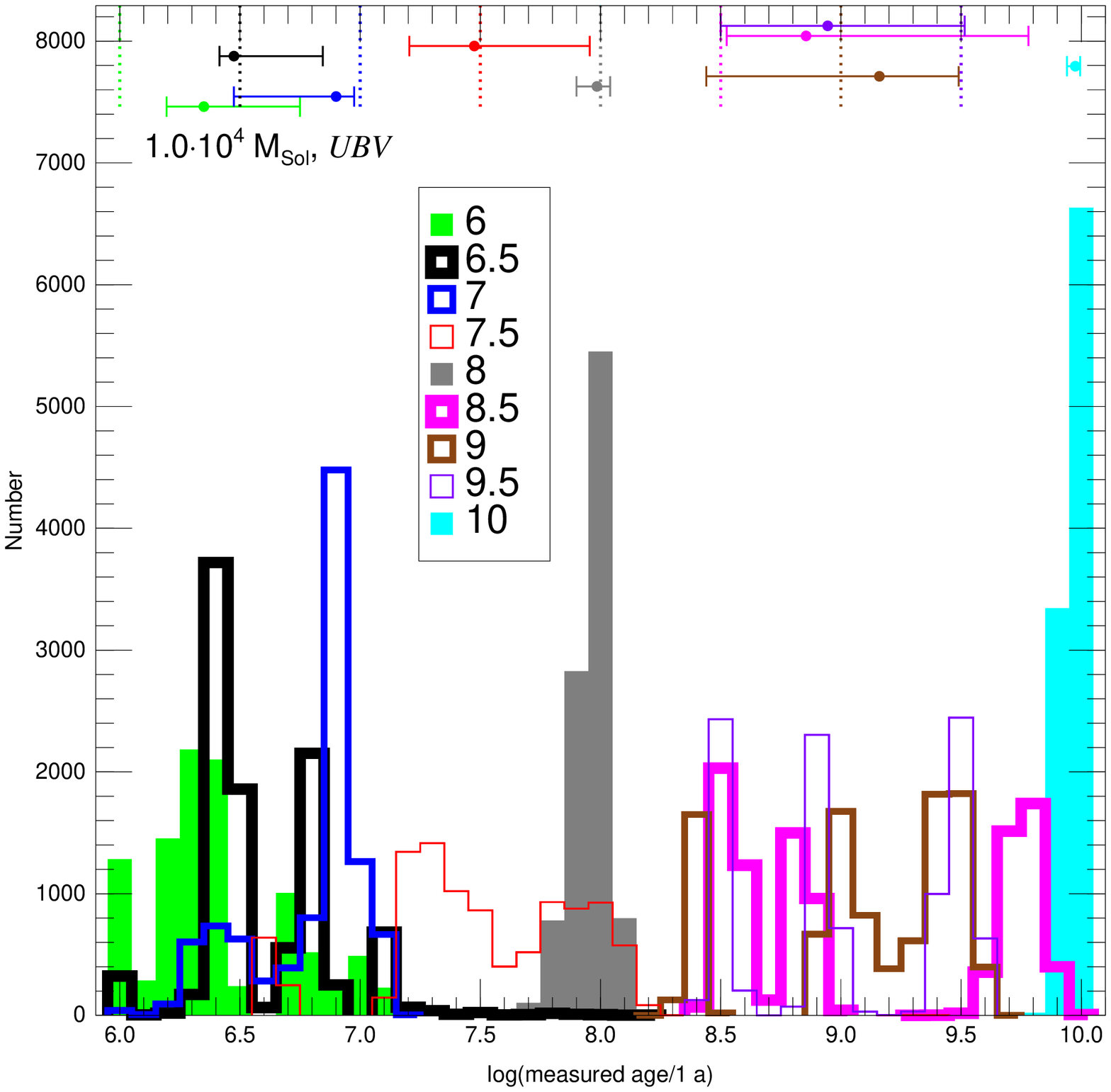}}
\centerline{\includegraphics*[width=0.42\linewidth]{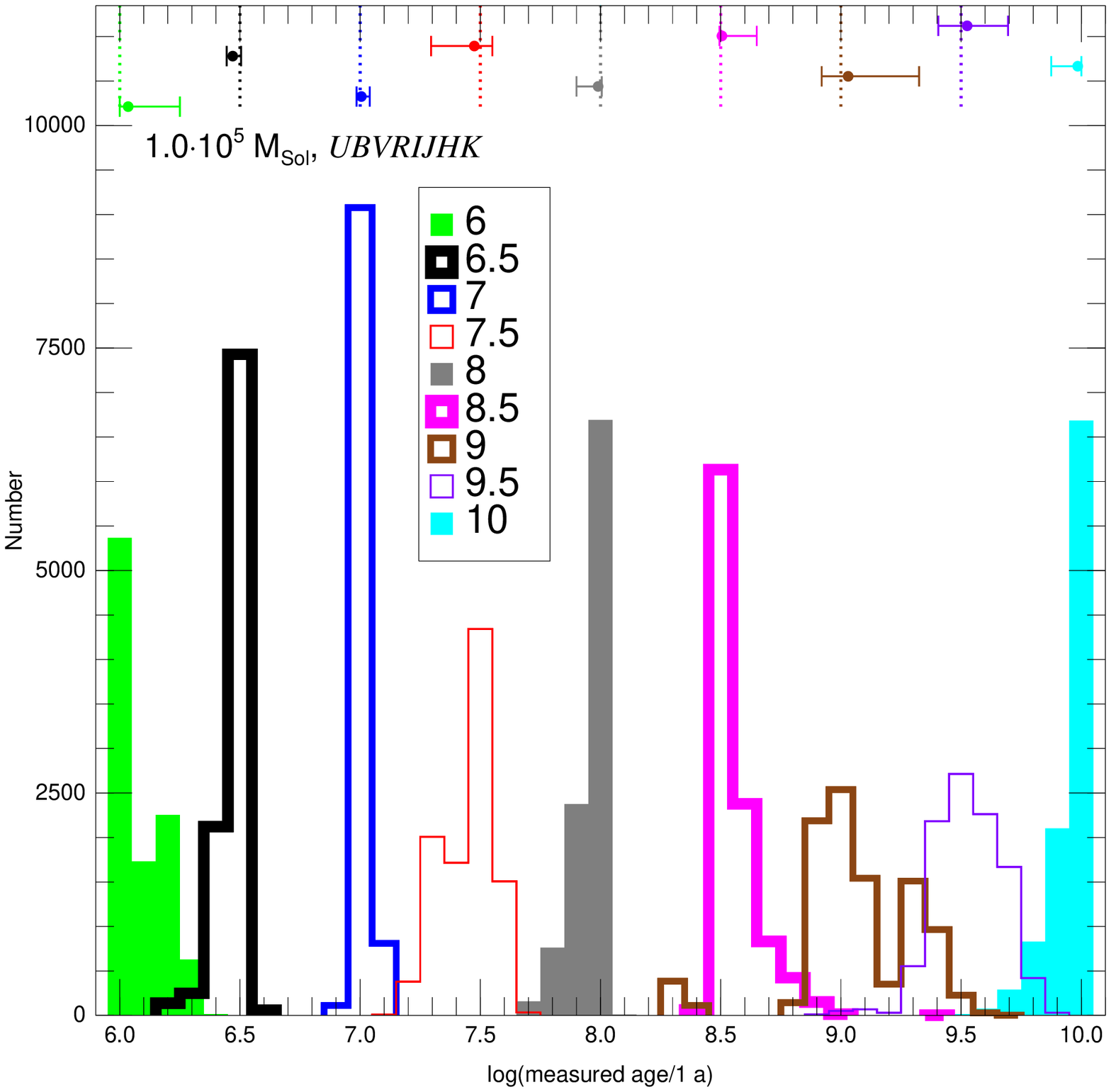}
            \includegraphics*[width=0.42\linewidth]{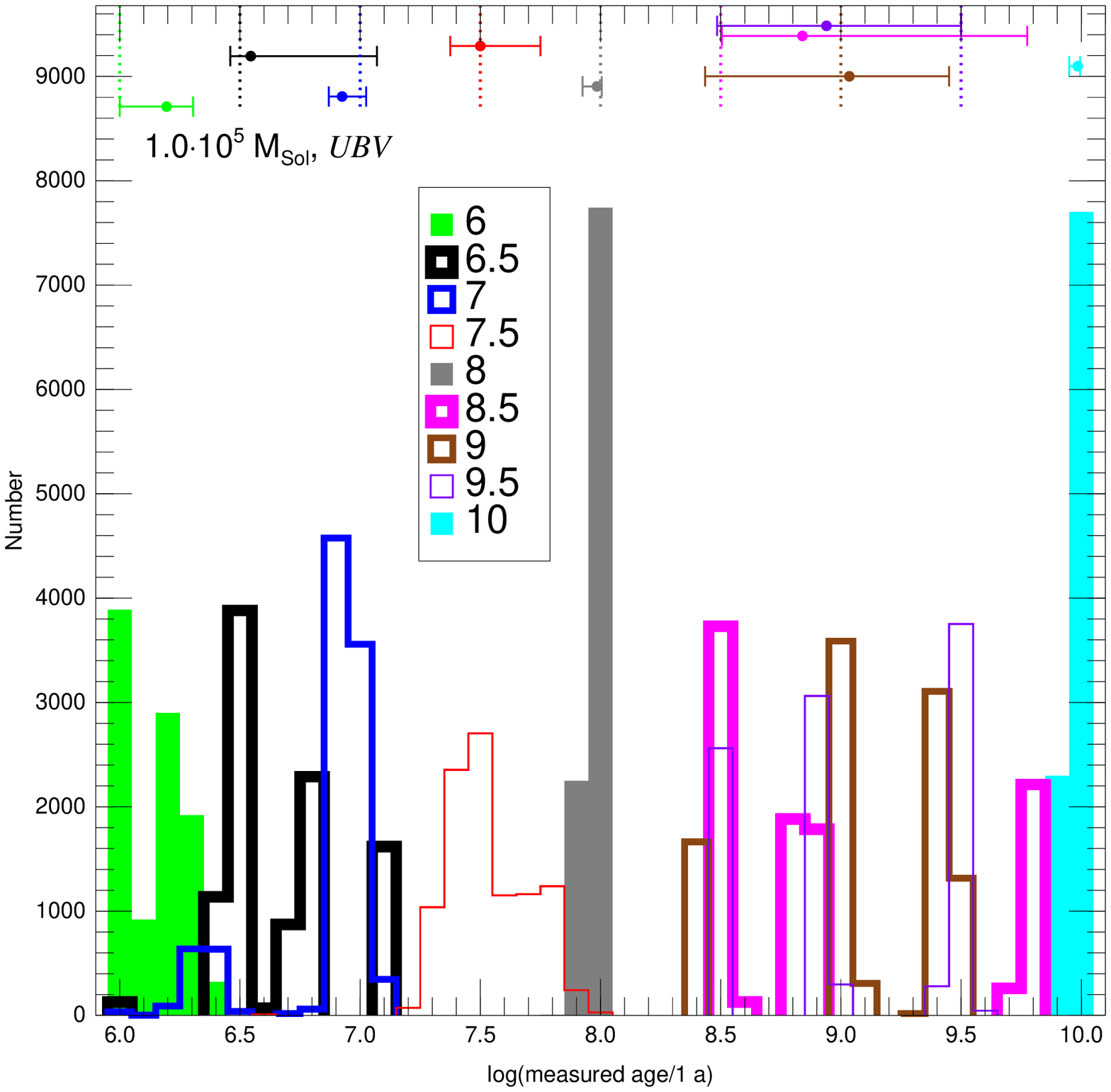}}
\caption{Distribution of observed ages for the nine input ages and two execution types (1: left plots, 
2: right plots) for clusters of $10^4$ M$_\odot$ (top) and $10^5$ M$_\odot$ (bottom) and unknown ages 
and extinctions. At the top of each panel the dashed line indicates the real age while the symbols and 
error bars provide the median and inferior and superior uncertainties (1-sigma equivalents) of each distribution.}
\label{agedist2}
\end{figure}	

	I used two execution types:

\begin{enumerate}
  \item $UBVRIJHK$ photometry with single-band $\theta$-dependent weights for each filter.
  \item $UBV$ photometry with single-band $\theta$-dependent weights for each filter.
\end{enumerate}

	The observed age distributions for $10^4$ M$_\odot$ and $10^5$ M$_\odot$ clusters are shown in 
Figure~\ref{agedist2}. The main results are:

\begin{itemize}
  \item As opposed to the cases where extinction is known, the addition of $RIJHK$ photometry provides a significant 
        improvement in the observed ages.
  \item $UBV$ photometry alone is not sufficient to accurately determine the ages of unresolved stellar clusters, even 
        for massive ones, if extinction is unknown.
  \item For some ages (1-10 Ma, 100 Ma, 10 Ga), $UBVRIJHK$ photometry provides relatively accurate observed ages for 
        $m = 10^4$ M$_\odot$. For the rest, higher masses are needed.
\end{itemize}

\section{Conclusions}

\begin{itemize}
  \item SIMF sampling effects can introduce large biases in the determination of masses and ages from unresolved 
        photometry.
  \item One should use weights based on the relative stochasticity of each filter when using 
        $\chi^2$-minimization techniques to calculate cluster properties from unresolved photometry.
  \item The addition of a broad-photometric baseline ($U$ to $K$) significantly reduces uncertainties when extinction is 
        determined from the data.
  \item For clusters younger than 100 Ma, a critical (post-RSG) stage exists around 30 Ma where ages are especially 
        difficult to determine.
\end{itemize}

%
%
%

\end{document}